\documentclass[a4paper]{jpconf}
\usepackage{graphicx}
\begin{document}
\title{Sensitivity of atmospheric neutrino experiments to neutrino non-standard interactions}

\author{Shinya Fukasawa}

\address{Department of Physics, Tokyo Metropolitan University, Minami-Osawa, Hachioji, Tokyo 192-0397, Japan}

\ead{fukasawa-shinya@ed.tmu.ac.jp}

\begin{abstract}
We study the sensitivity of atmospheric neutrino experiments to the neutrino non-standard interactions (NSI) which is motivated by the tension between the two mass squared differences extracted from the KamLAND and solar neutrino experiments. In this study the sensitivity of the future Hyper-Kamiokande experiments for 4438 days to NSI is shown. Assuming that the mass hierarchy is known,
we find that the best-fit value from the solar neutrino and KamLAND data
can be tested at more than 8 $\sigma$, while
the one from the global analysis can be examined at
5.0 $\sigma$ (1.4 $\sigma$) for the normal (inverted)
mass hierarchy.
\end{abstract}

\vspace*{-0.3cm}
\section{Introduction}
It is well established by solar, atmospheric, reactor and accelerator neutrino
experiments that neutrinos have masses and mixings.
In the standard three flavor neutrino oscillation framework,
there are three mixing angles $\theta_{12}$, $\theta_{13}$, $\theta_{23}$
and two mass-squared differences $\Delta m^2_{31}$, $\Delta m^2_{21}$.
However we do not know the value of the Dirac CP phase $\delta_{\rm CP}$,
the sign of $\Delta m^2_{31}$ (the mass hierarchy) and the octant
of $\theta_{23}$ (the sign of $\pi/4-\theta_{23}$).

We are entering the era of high precision neutrino oscillation oscillation measurements to measure the unknown parameters mentioned above.
Recently, it was pointed out that there is a tension in which smaller mass squared differences $\Delta m^2_{21}$ extracted from the global fit of the solar neutrinos are 2$\sigma$ smaller than that from the KamLand experiment.
The tension between KamLand and solar neutrinos can be solved by flavor-dependent nonstandard interactions (NSI).
Such a hint for NSI gives us a strong motivation to study NSI in propagation in details.

In Ref.\,\cite{Fukasawa:2015jaa} it was shown that the atmospheric neutrino measurements at Hyper-Kamiokande has a very good sensitivity to the NSI.
In this paper we discuss the sensitivity of the atmospheric neutrino measurements at Hyper-Kamiokande to NSI with a parametrization which is used in Ref.\,\cite{Gonzalez-Garcia:2013usa}.
The parametrization used in Ref.\,\cite{Gonzalez-Garcia:2013usa} is different from the ordinary one used in Ref.\,\cite{Fukasawa:2015jaa}.
Therefore a non-trivial mapping is required to compare the results in these two parametrization.
Our analysis was performed by taking this non-trivial mapping into account.

\vspace*{-0.3cm}
\section{The sensitivity of Hyper-Kamiokande to NSI}
In the three flavor
neutrino oscillation framework with NSI, the neutrino evolution is
governed by the Dirac equation:
\begin{eqnarray}
i {d \over dx} \left( \begin{array}{c} \nu_e(x) \\ \nu_{\mu}(x) \\ 
\nu_{\tau}(x)
\end{array} \right) = 
\left[  U {\rm diag} \left(0, \Delta E_{21}, \Delta E_{31}
\right)  U^{-1}
+{\cal A}\right]
\left( \begin{array}{c}
\nu_e(x) \\ \nu_{\mu}(x) \\ \nu_{\tau}(x)
\end{array} \right)\,,
\label{eqn:sch}
\end{eqnarray}
where
$\Delta E_{jk}\equiv\Delta m_{jk}^2/2E\equiv (m_j^2-m_k^2)/2E$,
$c_{jk}\equiv\cos\theta_{jk}$, $s_{jk}\equiv\sin\theta_{jk}$, $U$ stands for the leptonic mixing matrix and ${\cal A}$ stands for the matter potential with NSI expressed as
\begin{eqnarray}
{\cal A} \equiv
\sqrt{2} G_F N_e \left(
\begin{array}{ccc}
1+ \epsilon_{ee} & \epsilon_{e\mu} & \epsilon_{e\tau}\\
\epsilon_{\mu e} & \epsilon_{\mu\mu} & \epsilon_{\mu\tau}\\
\epsilon_{\tau e} & \epsilon_{\tau\mu} & \epsilon_{\tau\tau}
\end{array}
\right).
\label{matter-np}
\end{eqnarray}
In the investigation of the sensitivity of solar neutrinos to NSI \cite{Gonzalez-Garcia:2013usa}, a NSI parametrization $(\epsilon_D^f, \epsilon_N^f)$ which is different from that in equation (\ref{matter-np}) was introduced.
One can find the definition of $(\epsilon_D^f, \epsilon_N^f)$ in Ref.\,\cite{Gonzalez-Garcia:2013usa}.
Notice that the fermion subscript $f$ is not important in the case of atmospheric neutrinos with one particular choice of $f = u$
or $f = d$ at a time because the number densities of up and down quarks are approximately the same in the Earth.

We concentrate on only one particular choice of $f = d$ in this paper and therefore we omit the fermion subscript $f$ in our analysis.

Our aim is to investigate the sensitivity of the atmospheric neutrino experiment to NSI which is parametrized as ($\epsilon_{D}$, $\epsilon_{N}$) and this is done as follows.
\begin{enumerate}
  \item Set a grid on the ($\epsilon_{D}$, $|\epsilon_{N}|$) plane.
  \item Get a parameter set $\epsilon_{\alpha\beta}$ taking non-trivial mapping \cite{Fukasawa:2016nwn} into account for the given point ($\epsilon_{D}$, $|\epsilon_{N}|$) on the grid.
  \item Dismiss the parameter set if it is obviously excluded by the previous studies \cite{Friedland:2004ah,Friedland:2005vy,Fukasawa:2015jaa}.
  \item Calculate $\chi^2$ for each parameter set which passed the test (iii) and then obtain the minimum value of $\chi^2$ for the given ($\epsilon_{D}$, $|\epsilon_{N}|$).
\end{enumerate}

\begin{figure}[htbp]
\begin{minipage}{0.5\hsize}
\begin{center}
\includegraphics[width=14pc,angle=-90]{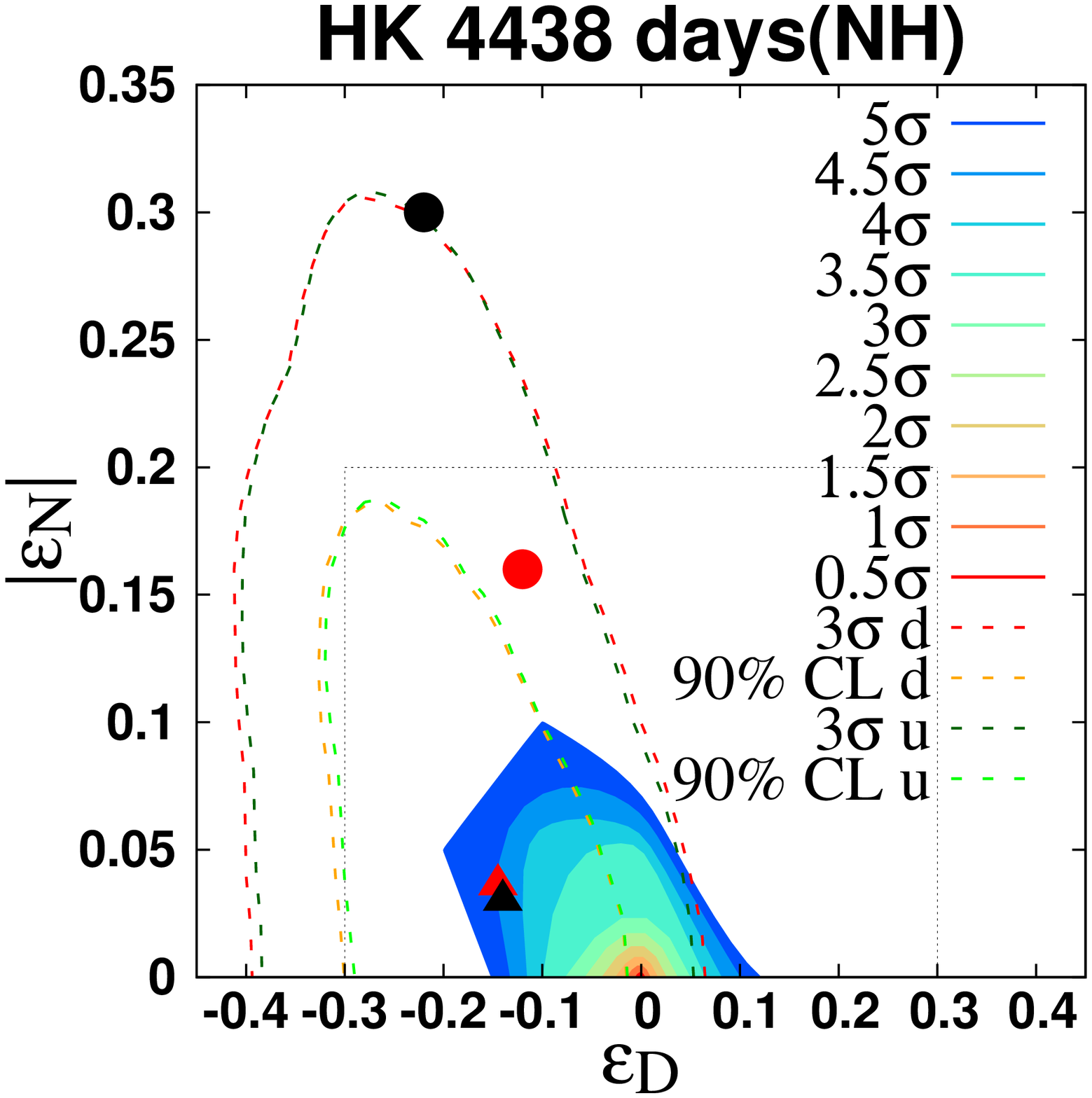}
\end{center}
\end{minipage}
\begin{minipage}{0.5\hsize}
\begin{center}
\includegraphics[width=14pc,angle=-90]{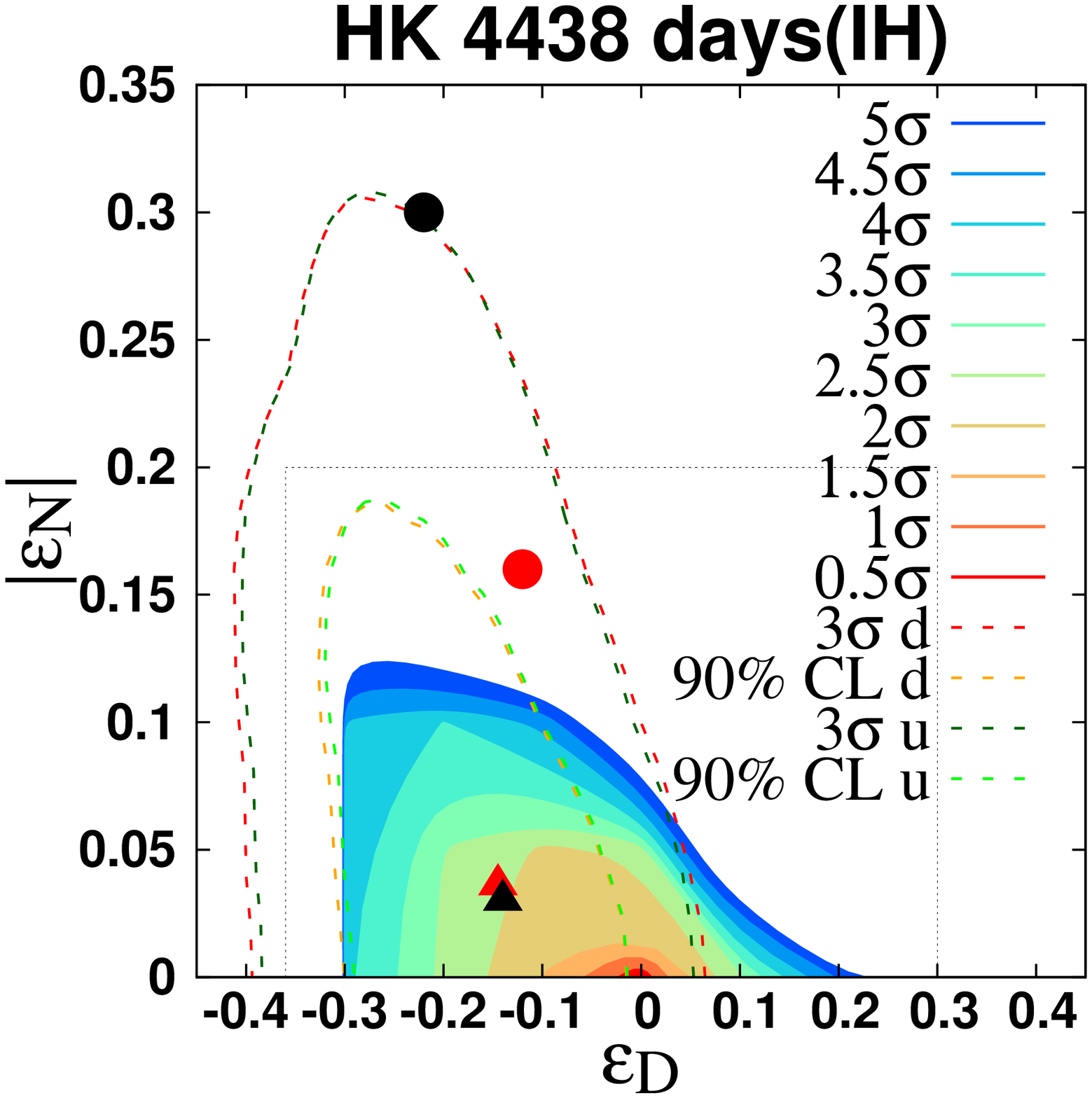}
\end{center}
\end{minipage}
\caption{The allowed region in the ($\epsilon_{D}$, $|\epsilon_{N}|$).
We calculated $\chi^2$ within the area surrounded by dotted lines and at the best fit points.
The triangles and circles are the best-fit values given by Ref.\,\cite{Gonzalez-Garcia:2013usa}.
The dashed lines are the boundaries of the allowed regions from the global neutrino oscillation experiments analysis.} 
\label{fig:fig1}
\vspace{0.5cm}
\begin{minipage}{0.5\hsize}
\begin{center}
\includegraphics[scale=0.25,angle=-90]{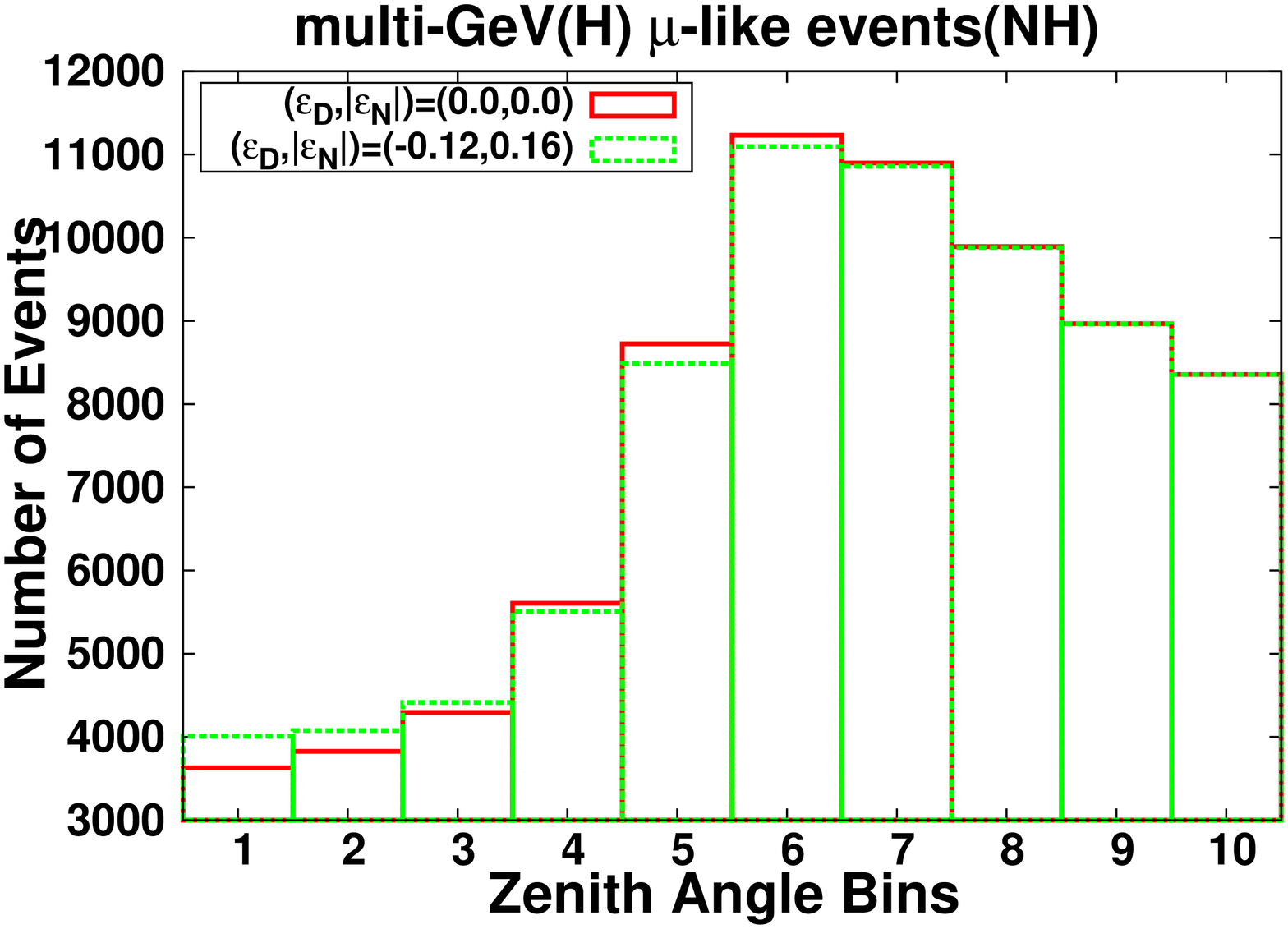}
\end{center}
\end{minipage}
\begin{minipage}{0.5\hsize}
\begin{center}
\includegraphics[scale=0.25,angle=-90]{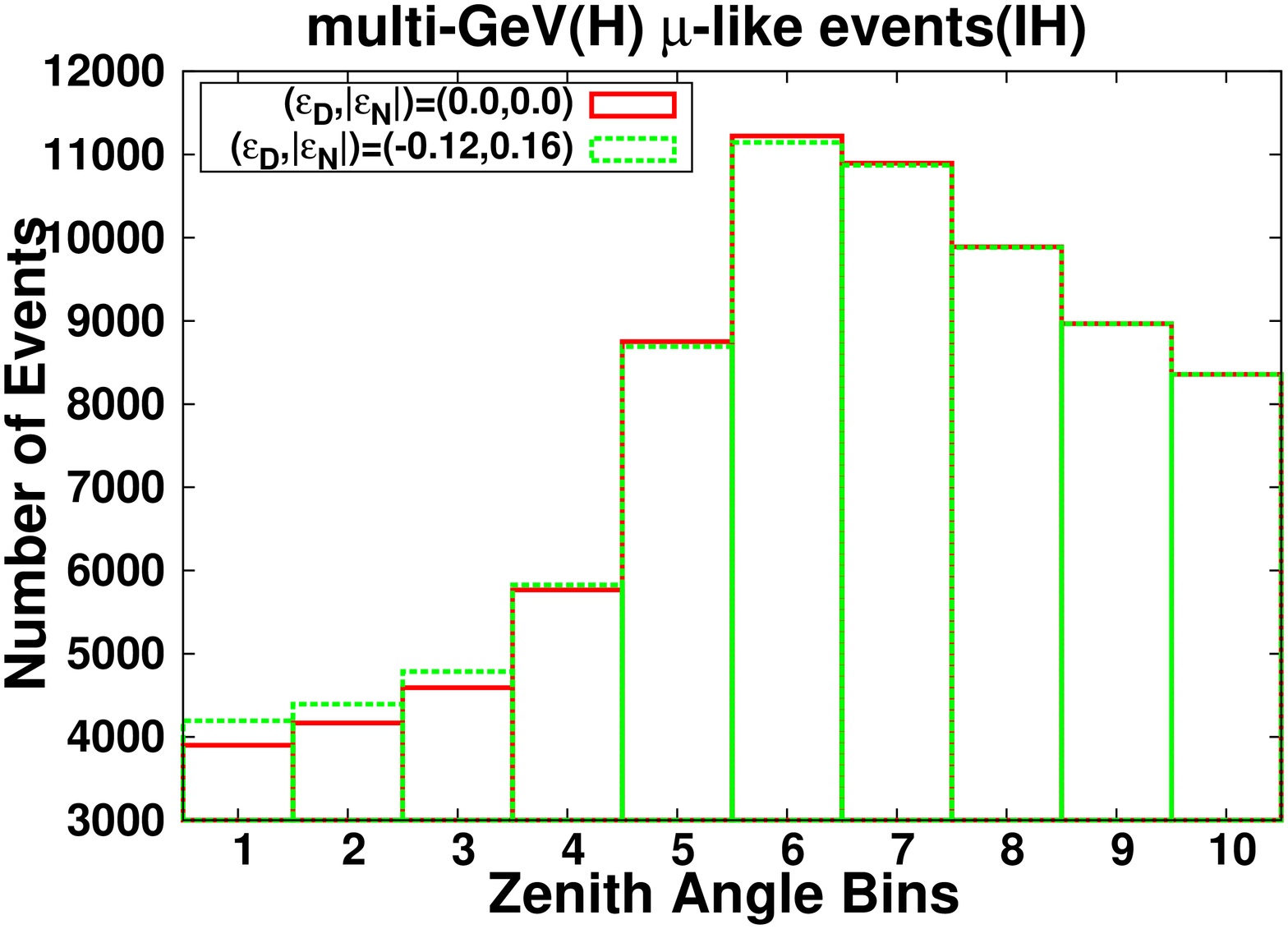}
\end{center}
\end{minipage}
\begin{minipage}{0.5\hsize}
\begin{center}
\includegraphics[scale=0.25,angle=-90]{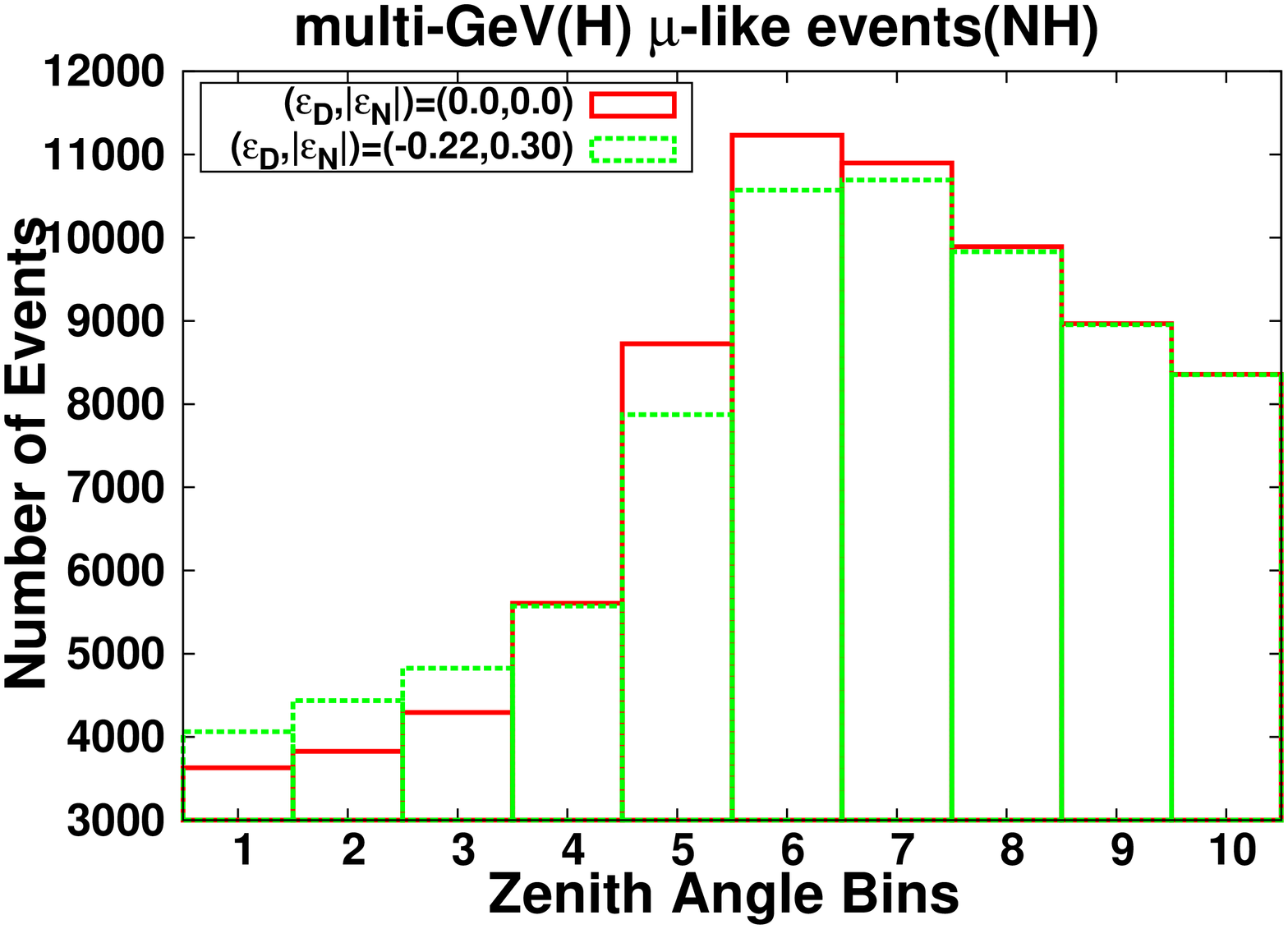}
\end{center}
\end{minipage}
\begin{minipage}{0.5\hsize}
\begin{center}
\includegraphics[scale=0.25,angle=-90]{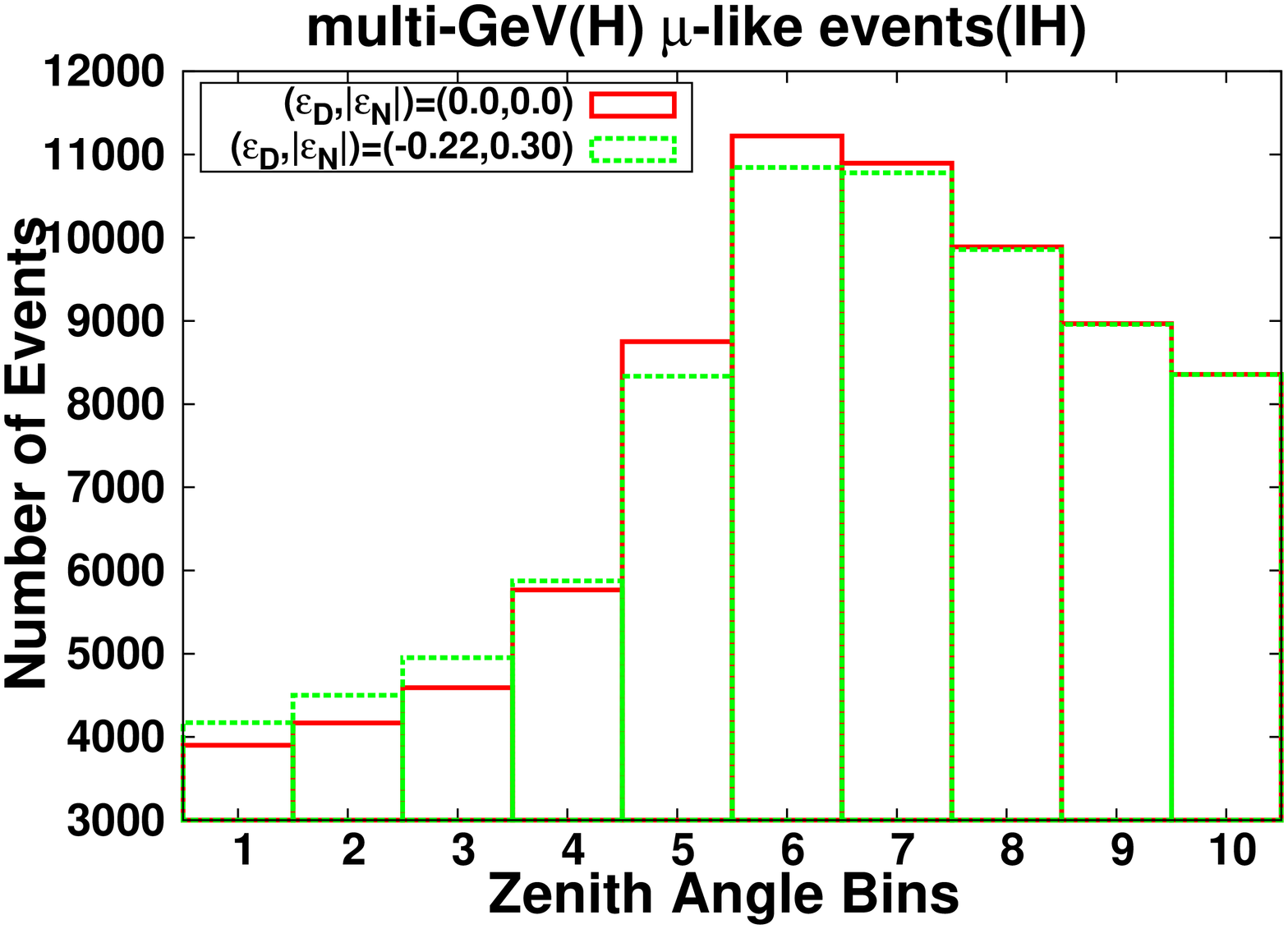}
\end{center}
\end{minipage}
\caption{The number of the high-energy-bin multi-GeV $\mu$-like events (red and green boxes are the standard scenario and the scenario with NSI, respectively). The horizontal axis is the zenith angle bin(1 for $-1.0<\cos\Theta<-0.8$, $\dots$, 10 for $0.8<\cos\Theta<1.0$ ). }
\label{fig:fig2}
\end{figure}

The results are shown in figure.\,\ref{fig:fig1}.
The best fit values $(\epsilon_{D}^d,\epsilon_{N}^d)=(-0.12,-0.16)$ for NSI with $f=d$ from the solar neutrino and KamLAND data given by Ref.\,\cite{Gonzalez-Garcia:2013usa} is excluded at $11\sigma$ ($8.2\sigma$) for the normal (inverted) hierarchy.
In the case of NSI with $f=u$, the best fit value $(\epsilon_{D}^u,\epsilon_{N}^u)=(-0.22,-0.30)$ is far from the standard scenario $(\epsilon_{D},\epsilon_{N})=(0.0,0.0)$ compared with the case of $f=u$ and also excluded at $38\sigma$ ($35\sigma$) for the normal (inverted) hierarchy.
On the other hand, the best fit value from the global analysis of the neutrino oscillation data \cite{Gonzalez-Garcia:2013usa} 
$(\epsilon_{D}^d,\epsilon_{N}^d)=(-0.145,-0.036)$ for NSI with $f=d$
is excluded at $5.0\sigma$ ($3.7\sigma$)  for the normal (inverted) hierarchy.
In the case of NSI with $f=u$, the best fit value 
$(\epsilon_{D}^u,\epsilon_{N}^u)=(-0.140,-0.030)$
is excluded at $5.0\sigma$ ($1.4\sigma$) 
for the normal (inverted) hierarchy.

To see which bin contributes to $\chi^2$ most, we focused on the difference between the number of events
of the standard scenario and that of the scenario with NSI (the red and black circle points in figure\,\ref{fig:fig1}).
Then we found that the multi-GeV $\mu$-like events coming from the below in the high-energy-bin contributes most to $\chi^2$.
This is because difference between the oscillation probability with NSI and without NSI is large where the neutrino energy is above 10 GeV.
We plotted the numbers of events for the multi-GeV $\mu$-like events in the high-energy-bin in figure\,\ref{fig:fig2}.

\vspace*{-0.3cm}
\section{Conclusion}
In this paper we have studied the sensitivity
of the future HK atmospheric neutrino
experiment to NSI which is suggested by
the tension between the mass squared differences
from the solar neutrino and KamLAND data.
We find that the best-fit value from the solar neutrino and KamLAND data
can be tested at more than 8 $\sigma$, while
the one from the global analysis can be examined at
5.0 $\sigma$ (1.4 $\sigma$) for the normal (inverted)
mass hierarchy.
It is worth noting that the scenario with NSI can be tested by looking at the multi-GeV $\mu$-like events in the future atmospheric neutrino experiments with high precision measurements.

\vspace*{-0.3cm}
\section*{Acknowledgments}
This research was partly supported by a Grant-in-Aid for Scientific
Research of the Ministry of Education, Science and Culture, under Grants No. 25105009, No. 15K05058, No. 25105001 and
No. 15K21734.

\end{document}